\documentclass[12pt]{article}
\usepackage{amsmath,amssymb}

\newcommand{\ch}{{\mathfrak h}}

\newcommand{\cQ}{{\cal Q}}
\newcommand{\cP}{{\cal P}}
\newcommand{\cR}{{\cal R}}

\newcommand{\cT}{{\cal T}}

\newcommand{\cB}{{\cal B}}
\newcommand{\cL}{{\cal L}}

\def\R{{\mathbb R}}

\newcommand{\lam}{\lambda}
\newcommand{\lan}{\langle}
\newcommand{\ran}{\rangle}
\newcommand{\beequ}{\begin{equation}}
\newcommand{\beeqn}{\begin{eqnarray}}
\newcommand{\eneqn}{\end{eqnarray}}

\newtheorem{thm}{Theorem}[section]

\newtheorem{prop}{Proposition}[section]
\newtheorem{lemm}{Lemma}[section]

\newtheorem{defn}{Definition}[section]
\newtheorem{rem}{Remark}[section]

\newenvironment{pf}{\bigskip\par\noindent{\it Proof:}}{$\square$\newline\vspace*{0.2cm}}

\makeatletter \@addtoreset{equation}{section}

\makeatother 
\title{{  Remarks on sufficient conditions for conservativity of minimal quantum dynamical semigroups \\[0.1cm]}}

\begin{document}
\newcommand\ds{\displaystyle}
\date{}
\maketitle
\baselineskip=18pt

\vspace{-2cm}
\begin{center}
\parbox{5in}
\noindent Changsoo Bahn

\noindent{\small  Natural Science Research Institute, Yonsei
University, Seoul 120-749, Korea \\[-0.1cm]
\small e-mail:bahn@yonsei.ac.kr}

\vskip 0.5 true cm

\noindent Chul Ki Ko

\noindent {\small Natural Science Research Institute, Yonsei
University,
Seoul 120-749, Korea\\[-0.1cm]
e-mail: kochulki@hotmail.com } \vskip 0.5 true cm

 \noindent Yong Moon Park

\noindent {\small Department of Mathematics, Yonsei University,
Seoul 120-749, Korea\\[-0.1cm]
e-mail: ympark@yonsei.ac.kr} \vskip 0.2 true cm
\end{center}

\begin{abstract}We obtain sufficient conditions for conservativity
of minimal quantum dynamical semigroup by modifying and extending
the method used in \cite{CF2}. Our criterion for conservativity
can be considered as a complement to Chebotarev and Fagnola's
conditions \cite{CF2}. In order to show that our conditions are
useful, we apply our results to a concrete example(  a model of
heavy ion collision).

\vspace*{0.3cm} \noindent {\it Keywords} : Quantum dynamical
semigroups; criterion for conservativity; Lindblad operators.
\end{abstract}

\baselineskip=24pt
\section{Introduction}
In this paper we are looking for any possible extension of
Chebotarev and Fagnola's sufficient conditions\cite{CF2} of
conservativity of minimal quantum dynamical semigroup. By
modifying  and extending the method employed in \cite{CF2}, we
obtain sufficient conditions for conservativity which extend the
previous one in some directions. In order to show that our
conditions are useful,  we apply our results to a  concrete
example (   a model of heavy ion collision).

The concept of quantum dynamical semigroup(q.d.s.) has become a
fundamental notion in study of irreversible evolutions in quantum
mechanics \cite{AL,BR}, open system \cite{Da} and quantum
probability theory [5 - 7]. 
  The theory of q.d.s. has
been intensively studied in recent years laying special emphasis
to the minimal q.d.s.  as well as to sufficient conditions to
ensure its conservativity (markovianity) [1, 8 - 13].  
It is worthy to mention that
there has been attention on the existence of stationary states for
a given conservative q.d.s. and faithfulness of the stationary
states\cite{FR1,FR2}.

 A q.d.s. $\cT = (\cT_t )_{t\ge 0}$ in $\cB( \ch)$, the Banach space of bounded operators in a
Hilbert space $\ch$, is a (ultraweakly continuous) semigroup of
completely positive linear maps on $\cB(\ch)$. A q.d.s. $\cT$ is
{\it conservative} if $\cT_t(I) = I$ where $I$ is the identity
operator on $\ch$.  In rather general cases, the infinitesimal
generator $\cL$ can be written (formally) as
\begin{equation}\label{1.1}
\cL(X) = i [ H , X] - \frac12 XM  + \sum_{l=1} ^\infty L_l ^* X
L_l - \frac 12 M X, \quad X \in B(\ch)
\end{equation}
where $M = \sum_{l=1}^\infty L_l ^* L_l,~ L_l$ is densely defined
and $H$  a symmetric operator on $\ch$\cite{Lin,Par}. However, for
unbounded generator $\cL$ in (\ref{1.1}) with (unbounded)
coefficients $H$ and $ L_l$, the solution $\cT$ of the quantum
master Markov equation
\begin{equation} \label{*1.2}
\frac d {dt} \cT_t (X) = \cL (\cT_t (X)), \quad \cT_0 (X) = X,
\end{equation}
may not be unique and conservative \cite{Ch1,BS}. Under suitable
conditions, the above equation (\ref{*1.2}) has a minimal solution
known as the minimal q.d.s.(see Sec. 2). Moreover if the minimal
q.d.s. is conservative, it is the unique solution of the above
equation. Also the study of conservativity conditions is important
in quantum probability because they play a key role in the proof
of uniqueness and unitarity of solutions of an
Hudson-Parthasarathy quantum stochastic differential equation
[18 - 20]. 

Chebotarev gave necessary and sufficient conditions for
conservativity \cite{Ch1}. Some of the conditions, however, are
impossible to check practically in many interesting examples.
Simplified forms of sufficient conditions were developed in
\cite{CF2,Ch2,CF1}. Especially the form of sufficient conditions
in \cite{CF2} can be written as follows: there exists a positive
self-adjoint operator $C$ bounded from below by $M$ satisfying a
form inequality
\begin{equation} \label{*1.3}
\cL (C) \le  b C
\end{equation}
where $b$ is a constant.

The main aim of this work is to improve the inequality
(\ref{*1.3}). Our form of sufficient conditions for conservativity
is as follows: there exists a positive self-adjoint operator $C$
bounded from below by $\delta M$ for some positive $\delta >0$
such that for all $\epsilon \in (0,1)$, two inequalities
\begin{eqnarray}\label{*1.4}
\cL (C)\quad \qquad &\le & \varepsilon  C^2 + b C + a \varepsilon^{-p}I ,\\
\label {*1.5} i [H,C] +  C^2 - \frac 12 (MC + CM)  &\le &
\varepsilon C^2 + b C + a \varepsilon^{-p}I
\end{eqnarray}
hold for some constants  $p \in (0,1)$, $ b \ge 0$ and $a \ge 0$.
 For details, see Theorem \ref{thm3.4}.

 In case the positive
self-adjoint operator $C$ satisfy (\ref{*1.5}), the inequality
(\ref{*1.4}) improves (\ref{*1.3}) obviously. Let us mention that
if we choose $M$ for $C$,
$$i[H,M] \le \varepsilon M^2 + b M + a \varepsilon^{-p}I$$
is  equivalent to (\ref{*1.5}). In order to explain our conditions
(\ref{*1.4}) and (\ref{*1.5}) are useful in practical sense, we
give some relative bounds(Lemma 4.1) and apply our result to a
concrete q.d.s. associated to a quantum system with dissipative
heavy ion collisions(Example 4.1). The conservativity of this
example has been already considered in \cite{CF2}. However,
applying our criterion, we are able to control local singularities
of (derivatives of ) coefficients of the infinitesimal generator
(see Remark \ref{rem4.2}).

 The paper is organized as follows. In Sec. 2, we give a brief review
on the theory of minimal q.d.s. and criteria for conservativity.
In Sec. 3, we first list  our sufficient conditions for
conservativity and then produce the proof of our result. In Sec.
4, we give some relative bounds to apply the results of Sec. 3 to
a concrete q.d.s..

\section{The minimal quantum dynamical semigroup}
\quad Let $\ch$ be a complex separable Hilbert space with the
scalar product $\langle\cdot,\cdot\rangle$ and norm $\|\cdot\|$.
Let $\mathcal{B}(\ch)$ denote the Banach space of bounded linear
operators on $\ch$. The uniform norm in $\mathcal{B}(\ch)$ is
denoted by $\|\cdot\|_\infty$ and the identity in $\ch$ is denoted
by $I$. We denote by $D(G)$ the domain of operator $G$ in $\ch$.
\begin{defn}
A quantum dynamical semigroup (q.d.s.) on $\mathcal{B}(\ch)$ is a
family $\mathcal{T}=(\mathcal{T}_t)_{t\ge0}$ of operators in
$\mathcal{B}(\ch)$ with the following properties:
\begin{enumerate}
\item[(i)] $\mathcal{T}_0(X)=X,$ for all $X\in\mathcal{B}(\ch),$
\item[(ii)]$\mathcal{T}_{t+s}(X)=\mathcal{T}_{t}(\mathcal{T}_{s}(X)),$
for all $s,t\ge0$ and all $X\in\mathcal{B}(\ch),$
\item[(iii)]$\mathcal{T}_t(I)\le I,$ for all $t\ge0,$
\item[(iv)](completely positivity) for all $t\ge0$, all integer
$n$ and all finite sequences $(X_j)_{j=1}^n,\,(Y_l)_{l=1}^n$ of
elements of $\mathcal{B}(\ch)$, we have $$\sum_{j,\,l=1}^n
Y_l^*\mathcal{T}_t(X_l^*X_j)Y_j\ge0,$$ \item[(v)] (normality) for
every sequence $(X_n)_{n\ge1}$ of $\mathcal{B}(\ch)$ converging
weakly to an element $X$ of $\mathcal{B}(\ch)$ the sequence
($\mathcal{T}_t(X_n))_{n\ge1}$ converges weakly to an element
$\mathcal{T}_t(X)$ for all $t\ge0$,
 \item[(vi)] (ultraweak continuity) for
all trace class operator $\rho$ on $\ch$ and all
$X\in\mathcal{B}(\ch)$ we have
$$\lim_{t\rightarrow0^+}
Tr(\rho\mathcal{T}_t(X))=Tr(\rho X).$$
\end{enumerate}
\end{defn}
We recall that  as a consequence of properties (iii), (iv)  for
each  $t\ge0$ and $X \in \cB(\ch)$, $ \cT_t$ is
 a contraction, i.e.,
 \begin{equation}\label{2.0}
 \| \mathcal{T}_{t}(X)\|_{\infty}\le\|X\|_{\infty}.
 \end{equation}
 Also recall that as a consequence of properties (iv), (vi), for all $X\in\mathcal{B}(\ch)$, the map
  $t\mapsto \mathcal{T}_{t}(X)$ is strongly continuous.

  \begin{defn}
A q.d.s. $\mathcal{T}=(\mathcal{T}_t)_{t\ge0}$ on
$\mathcal{B}(\ch)$
 is called to be
 conservative  if
$\mathcal{T}_{t}(I)=I$ for all $t\ge0$.
\end{defn}

As mentioned in Introduction, the natural generator of q.d.s.
would be the Lindblad type generator\cite{Lin,Par}. Letting
\begin{equation} \label{*2.2}
G=-iH -\frac12 M,   \quad \text{where} \,\, M =\sum_{l=1}^\infty
L_l^* L_l,
\end{equation}
 the  infinitesimal generator in (\ref{1.1}) can
be formally written by
$$ \cL(X) =XG + G^* X + \sum_{l=1}^\infty
L_l^* X L_l. $$ A very large class of q.d.s. was constructed by
Davies \cite{Da2} under the following assumption.

\vspace{0.5cm} \noindent {\it  {\bf A.} The operator $G$ is the
infinitesimal generator of a strongly continuous contraction
semigroup $P=(P(t))_{t\ge 0} $ in $\ch$.  The domain of the
operators $(L_l )_{l=1} ^\infty $ contains the domain $D(G)$
 of the operator $G$. For all $v, u \in D(G)$, we have
\begin{equation}\label{2.1}
\langle v,Gu \rangle+\langle Gv,u \rangle+\sum_{l=1}^\infty \langle L_l v,L_l u\rangle=0.
\end{equation}
}

 As a result of Proposition 2.5 of \cite{CF1}, we can assume only that
the domain of the operators $L_l$ contains a subspace $D$ which is
a core for $G$ and (\ref{2.1}) holds for all $v,u\in D$.

For all $X\in \mathcal{B}(\ch)$, consider the sesquilinear form
$\mathcal{L}(X)$ on $\ch$ with domain $D(G)\times D(G)$ given by
\begin{equation}\label{2.2}
\langle v,\mathcal{L}(X)u\rangle=\langle v,XGu \rangle+\langle
Gv,Xu \rangle+\sum_{l=1}^\infty\langle L_l v,XL_l u\rangle.
\end{equation}
 Under the assumption {\bf A}, one can construct a q.d.s.
$\cT=(\cT_t)_{t\ge0}$ satisfying the equation
 \begin{equation}\label{2.3}
\langle v,\cT_t(X)u \rangle=\langle v,Xu \rangle+\int_0^t\langle
v,\mathcal{L}(\cT_s(X))u\rangle ds
\end{equation}
 for all $v ,u\in D(G)$ and all $X\in B(\ch).$ For a strongly
continuous family $(\cT_t(X))_{t\ge0}$ of elements of
$\mathcal{B}(\ch)$ satisfying (\ref{2.0}), the followings are
equivalent:
\begin{enumerate}
\item[(i)] equation (\ref{2.3}) holds for all $v ,u\in D(G)$,
\item[(ii)] for all $v ,u\in D(G)$ we have
\begin{eqnarray}\label{2.4}
\langle v,\cT_t(X)u \rangle&=&\langle P(t) v, X P(t) u
\rangle\nonumber\\&&+\sum_{l=1}^\infty\int_0^t\langle L_lP(t-s)
v,\cT_s(X)L_l P(t-s) u\rangle ds.
\end{eqnarray}
\end{enumerate}
We refer to the proof of Proposition 2.3 in \cite{CF2}.
 A solution of the equation (\ref{2.4}) is
obtained by the iterations
\begin{eqnarray}\label{2.5}
\langle u,\cT_t^{(0)}(X)u\rangle&:=&\langle
P(t)u,X P(t)u\rangle,\nonumber\\
\langle u,\cT_t^{(n+1)}(X)u\rangle&:=&\langle P(t)u,X P(t)u\rangle\nonumber\\
&&+\sum_{l=1}^\infty\int_0^t \langle
L_lP(t-s)u,\cT_s^{(n)}(X)L_lP(t-s)u\rangle ds
\end{eqnarray}
for all $u\in D(G)$. In fact, for all positive elements
$X\in\mathcal{B}(\ch)$ and all  $t\ge0$, the sequence of operators
$(\cT_t^{(n)}(X))_{n\ge0}$ is non-decreasing. Therefore it is
strongly convergent and  its limits for $X\in\mathcal{B}(\ch)$ and
$t\ge0$ define the {\it minimal solution} $\cT^{(min)}$ of
(\ref{2.4}) in the sense that, given another solution
$(\cT_t')_{t\ge0}$ of (\ref{2.3}), one can easily check that
$$\cT^{(min)}_t(X)\le \cT_t'(X)\le \|X\|_\infty  I$$ for any
positive element $X$ and all $t\ge0$. For details, we refer to
\cite{Ch1,Fa1}.

We recall here a necessary and sufficient condition for
conservativity of minimal q.d.s. obtained by Chebotarev. Let us
consider the linear monotone maps $\cP_\lam : \mathcal{B}(\ch) \to
\mathcal{B}(\ch)$ and $\cQ_\lam : \mathcal{B}(\ch) \to
\mathcal{B}(\ch)$ defined by
\begin{equation} \label{2.6}
\lan v, \cP_\lam (X) u\ran = \int^\infty_0 e^{-\lam s} \lan P(s)
v, X P(s) u \ran ds,
\end{equation}
\begin{equation} \label{2.7}
\lan v, \cQ_\lam (X) u\ran =\sum_{l=1}^\infty \int^\infty_0
e^{-\lam s} \lan L_l P(s) v, X L_l P(s) u \ran ds
\end{equation}
for all $\lam >0$ and $X \in \mathcal{B}(\ch), \, v,u \in D(G)$.
It is easy to check that both $\cP_\lam$ and $\cQ_\lam$ are
completely positive, and also both $\lam\cP_\lam$ and $\cQ_\lam$
are normal contractions in $\mathcal{B}(\ch)$ (see Sec. 2 of
\cite{CF1}).

The resolvent of the minimal q.d.s. $(\cR^{(min)}_\lam )_{\lam
>0}$ defined by
$$ \lan v, \cR^{(min)}_\lam (X) u\ran = \int^\infty_0 e^{-\lam s}
\lan  v, \cT^{(min)}_s(X) u \ran ds $$ (with $X \in B(\ch)$ and
$v,u \in \ch$ ) can be represented as
\begin{equation} \label{1.7-1}
\cR^{(min)}_\lam (X) = \sum_{k=0}^\infty \cQ^k_\lam (\cP_\lam
(X)),
\end{equation}
the series being convergent for the strong operator topology(see
Theorem 3.1 of \cite{CF2}).

\begin{prop}\label{prop2.1} Suppose that
the condition {\bf A} holds and fix $\lam >0$. Then the sequence
of positive operators $(\cQ^k _\lam (I))_{k \ge 0}$ is
non-increasing. Moreover the following conditions are equivalent:
\begin{enumerate}
\item[(i)] the minimal q.d.s. $\cT^{(min)}$ is conservative,
 \item[(ii)]  s-$\lim_{k \to \infty} \cQ^k_\lam (I) = 0$.
\end{enumerate}
\end{prop}
The above proposition has been proved in \cite{CF2,CF1}. Due to
Proposition \ref{prop2.1}, the minimal q.d.s. is conservative
whenever, for a fixed $\lam
>0$, the series
\begin{equation} \label{3.1}
\sum_{k=0}^\infty \lan u, \cQ^k_\lam (I) u\ran
\end{equation}
is convergent for all $u$ in a dense subspace of $\ch$. In fact in
this case, the condition (ii) of Proposition \ref{prop2.1} holds
because the sequence of positive operators
$(\cQ^k_\lam(I))_{k\ge0}$ is non-increasing.

Employing the above facts, Chebotarev  and Fagnola have obtained a
criteria to verify the conservativity of minimal q.d.s. ( see Sec.
4 in \cite{CF2}). Here we give their result(Theorem 4.4 in
\cite{CF2}):
\begin{thm}\label{thm2.2} Under the assumption {\bf A} suppose
that there exists a positive self-adjoint operator $C$ in $\ch$
with the following properties:
\begin{enumerate}
\item[(a)] the domain $D(G)$ of $G$ is contained in the domain of
the positive square root $C^{1/2}$ and $D(G)$ is a core for
$C^{1/2}$ ,
 \item[(b)] the linear manifolds $L_l (D(G^2))$, $l \ge 1,$ are contained in the
domain of $C^{1/2}$,
\item [(c)] there exists a self-adjoint
operator $\Phi$, with $D(G)\subset D(\Phi^{1/2})$ and $D(C)\subset
D(\Phi)$, such that, for all $u\in D(G)$, we have
$$-2 Re \lan u,Gu \ran  =\sum_{l=1}^\infty \|L_l u\|^2 =\|\Phi^{1/2}u\|^2,$$
 \item[(d)] for all $u\in D(C)$ we have $\|\Phi^{1/2}u\|\le\|C^{1/2}u\|,$
\item[(e)] for all $u\in D(G^2)$ there exists a positive constant $b$ depending only on $G,\,C,\,L_l$
\begin{equation} \label{1.8}
2 Re \lan C^{1/2} u, C^{1/2} G u \ran + \sum_{l=1}^\infty
\|C^{1/2}L_l u\|^2\le b \| C^{1/2} u \|^2.
\end{equation}
\end{enumerate}
Then the minimal q.d.s. is conservative.
\end{thm}
We will call the conditions in Theorem \ref{thm2.2} {\it C-F
sufficient condition}.

\section{Sufficient condition for conservativity}
\quad In this section we extend more or less C-F sufficient
condition for conservativity of the minimal q.d.s.. First  we
introduce our assumption.

\vspace{0.5cm}
 \noindent  {\bf C}. {\it There exists a positive
self-adjoint operator $C$ such that
\begin{description}
\item[(a)] the domain of its positive square root $C^{1/2}$
contains the domain $D(G)$ of $G$ and $D(G)$ is a  core of
$C^{1/2}$. Also the domain of $C$ contains the domain of $G^2$.
 \item[(b)] the linear manifolds $L_l (D(G^2))$, $l
\ge 1$, are contained in the domain of $C^{1/2}$,
 \item [(c)] there exist $p\in(0,\,1)$, $b\ge0$ and $a\ge0$ such that for any
$\varepsilon\in (0,\,1)$ two inequalities
\begin{equation} \label{2.12}
2 Re \lan C u,  G u \ran  \le -(1-\varepsilon) \| Cu\|^2 + b \|
C^{1/2} u \|^2 + a \varepsilon^{-p} \|u \|^2
 \end{equation}
 and
\begin{eqnarray} \label{2.13}
&&2 Re \lan C u,  G u \ran + \sum_{l=1}^\infty \|C^{1/2}L_l u\|^2
\nonumber\\ &&\hspace{3cm}\le \varepsilon \| Cu\|^2 + b \| C^{1/2}
u \|^2 + a \varepsilon^{-p} \|u \|^2
\end{eqnarray}
hold for all $u \in D(G^2)$.
\end{description}
}

The following is our main result:
\begin{thm} \label{thm3.4}
Suppose that assumptions {\bf A} and  {\bf C} hold for some
positive self-adjoint operator $C$ and there exists a positive
self-adjoint operator $\Phi$ in $\ch$ such that:
\begin{enumerate}
\item [(a)]the domain of the positive square root $\Phi^{1/2}$
contains the domain of $G$ and, for every $u \in D(G)$, we have
$$
-2 Re \lan u, Gu \ran = \sum_{l=1}^\infty \lan L_l u , L_l u\ran =
\lan \Phi ^{1/2} u , \Phi^{1/2} u \ran,
$$
\item[(b)] the domain of $C$ is contained in the domain $\Phi$
and, for some $\delta>0$, we have
$$ \delta\lan \Phi^{1/2} u , \Phi^{1/2} u \ran \le \lan C^{1/2} u,
C^{1/2} u \ran,\quad \forall u \in D(C). $$
\end{enumerate}
Then the minimal q.d.s. is conservative.
\end{thm}
Before proceeding the proof of the above theorem, it may be worth
to give some remarks on the assumption {\bf C}.
\begin{rem}\label{rem2.3}  (a) If we choose the operator $C$ satisfying (\ref{2.12}), the
inequality (\ref{2.13})  evidently improves (\ref{1.8}) in C-F
sufficient condition.

(b) As mentioned in Introduction, the inequality (\ref{2.12}) can
be written formally by
$$
i [H,C] +  C^2 - \frac 12 (MC + CM)  \le \varepsilon C^2 + b C + a
\varepsilon^{-p}I.
$$
 If we choose $ C=M(=\sum_{l=1}^\infty L_l^*
L_l)$,  then  (\ref{2.12})  is equivalent to the following
condition $$ i \lan u,[H,M]u \ran \le \varepsilon \|M u\|^2 + b
\|M^{1/2} u\|^2 + a \varepsilon^{-p}\| u \|^2. $$ Thus, in many
cases the condition (\ref{2.12}) is easier to check than
(\ref{2.13}).

(c) As Kato's relative bounds\cite{Ka} control local singularities
of potentials in the Schr\"odinger operator, we believe that the
bounds in (\ref{2.12}) and (\ref{2.13}) will be able to control
local singularities of (derivatives of) the coefficients of
generators of q.d.s..
\end{rem}

 In the rest of
this section we produce the proof of Theorem \ref{thm3.4}. The
following is an extension of the condition that the series
(\ref{3.1}) converges.

\begin{lemm}\label{lemm2.1}  Suppose that for fixed $\lambda>0$, the series
\begin{equation} \label{3.2}
\sum_{k=0}^\infty \frac1{k+1} \lan u, \cQ^k_\lam (I) u\ran
\end{equation}
is convergent for all $u$ in a dense subspace of $\ch$. Then we
have s-$\lim_{k \to \infty} \cQ^k_\lam (I) = 0$.
\end{lemm}
\begin{pf}
Notice that $(\cQ^k_\lam(I))_{k\ge 0}$ is a positive and
non-increasing sequence. Therefore it is strongly convergent to a
positive operator $Y$, i.e., $$Y:=s-\lim_{k \to \infty} \cQ^k_\lam
(I)\ge0.$$
 Suppose that $Y$ is not zero. Then there exists a
non-zero vector $u \in \ch$ such that $\langle u,Yu\rangle>0$.
This implies that
$$0\,<\,\langle u,Yu\rangle\,\le\, \langle u, \cQ^k_\lam
(I)u\rangle\quad \text{for\,\,all}\,\,k\ge0,$$ and also
$$\langle u,Yu\rangle\sum_{k=0}^{n}\frac{1}{k+1}\,\le\,
\sum_{k=0}^{n}\frac{1}{k+1}\langle u, \cQ^k_\lam (I)u\rangle$$ for
any nonnegative integer $n$. Thus the series (\ref{3.2}) is
divergent, which is contrary to the assumption. Thus $Y$ must be
zero.
\end{pf}

\noindent By Proposition \ref{prop2.1} and Lemma \ref{lemm2.1},
the minimal q.d.s. is conservative whenever, for a fixed $\lam
>0$, the series
\begin{equation*}
\sum_{k=0}^\infty \frac1{k+1} \lan u, \cQ^k_\lam (I) u\ran
\end{equation*}
converges for all $u$ in a dense subspace of $\ch$.
 By Monotone Convergence Theorem, we have
\begin{equation}\label{3.3}
\sum_{k=0}^\infty \frac1{k+1} \lan u, \cQ^k_\lam (I)u\ran
=\int_0^1 \Big(\sum_{k=0}^\infty x^k \lan u, \cQ^k _\lam (I) u
\ran\Big) dx.
\end{equation}

Fix $x\in(0,\,1).$ For all $u\in D(G)$ and $X\in\mathcal{B}(\ch)$,
let $\cT^{(min)}_{t,x}(X)$ be the solution obtained by the
iterations
\begin{eqnarray}\label{3.4}
\langle u,\cT_{t,x}^{(0)}(X)u\rangle&=&\langle
P(t)u,XP(t)u\rangle,\nonumber\\\langle
u,\cT_{t,x}^{(n+1)}(X)u\rangle&=&\langle P(t)
u,XP(t)u\rangle\nonumber\\&&+x\sum_{l=1}^\infty\int_0^t \langle
L_l P(t-s) u,\cT_{s,x}^{(n)}(X)L_l P(t-s)u\rangle ds.
\end{eqnarray}
For all $u\in \ch$ and $X\in\mathcal{B}(\ch)$, and for
$\lambda>0$, let
\begin{eqnarray}\label{3.5}
\langle u,\cR_{\lambda,x}^{(n)}(X)u\rangle&=&\int_0^\infty
e^{-\lambda t} \langle u,\cT_{t,x}^{(n)}(X)u\rangle dt,\\
\langle u,\cR_{\lambda,x}^{(min)}(X)u\rangle &=&\int_0^\infty
e^{-\lambda t} \langle u,\cT_{t,x}^{(min)}(X)u\rangle dt.\nonumber
\end{eqnarray}
Clearly (\ref{2.0}) guarantees that $\cR_{\lambda,x}^{(n)}(X)$ and
$\cR_{\lambda,x}^{(min)}(X)$ are well defined. We can also obtain
the relation corresponding to (\ref{1.7-1}).
\begin{prop}
For any $x \in(0,\,1), \lambda>0$ and $X\in\mathcal{B}(\ch)$ we
have
\begin{equation}\label{3.5-1}
\cR_{\lam, x}^{(min)} (X) = \sum_{k=0}^\infty x^k \cQ_\lam ^k
(\cP_\lam (X))
\end{equation}
the series being convergent for the strong operator
topology.
\end{prop}
\begin{pf} For any positive element $X$ of $\mathcal{B}(\ch)$,
the sequence $(\cR_{\lambda, x}^{(n)}(X))_{n\ge0}$ is
non-decreasing. Therefore by (\ref{3.5}), for all $u\in\ch$ we
have
\begin{eqnarray*}
\langle u,\cR_{\lambda,x}^{(min)}(X)u\rangle&=&\int_0^\infty
e^{-\lambda t} \langle u,\cT_{t,x}^{(min)}(X)u\rangle
dt\\&=&\sup_{n\ge0}\langle u,\cR_{\lambda,x}^{(n)}(X)u\rangle.
\end{eqnarray*}
The second equation (\ref{3.4}) yields
\begin{eqnarray}\label{3.6}
\langle u,\cR_{\lambda,x}^{(n+1)}(X)u\rangle &=&\int_0^\infty
e^{-\lambda t} \langle P(t)u, XP(t)u\rangle dt\\
       &&+x\sum_{l=1}^\infty\int_0^\infty
e^{-\lambda t} \int_0^t\langle L_lP(t-s) u,\cT_{s,x}^{(n)}(X)L_l
P(t-s) u\rangle ds dt\nonumber
\end{eqnarray}
for all $u\in D(G)$. By the change of variables in the above
double integral and (\ref{2.6}) we have
\begin{eqnarray}\label{3.6-0}
\langle u,\cR_{\lambda,x}^{(n+1)}(X)u\rangle &=&\langle u, \cP_{\lambda}(X)u\rangle \\
       &&+x\sum_{l=1}^\infty\int_0^\infty
e^{-\lambda r} \int_0^\infty e^{-\lambda s}\langle L_l
P(r)u,\cT_{s,x}^{(n)}(X)L_l P(r) u\rangle ds dr.\nonumber
\end{eqnarray}
Thus we obtain the recursion relation
 $$\cR^{(n+1)}_{\lam , x}
(X)= \cP_\lam (X) + x \cQ_\lam (\cR^{(n)}_{\lam , x} (X)).
$$
Iterating $n$ times, we have
\begin{equation} \label{2.11}
\cR^{(n+1)}_{\lam , x} (X)= \sum_{k=0} ^{n+1} x^k \cQ^k_\lam
(\cP_\lam (X))
\end{equation}
and (\ref{3.5-1}) follows from letting $n$ tend to $\infty$. Since
any bounded operator can be written as a linear combination of
four positive self-adjoint operators (\ref{3.5-1}) also holds for
an arbitrary element of $\mathcal{B}(\ch)$.
\end{pf}

\begin{lemm} \label{rem2.4}
 Condition {\bf C} implies that, for each $u \in D(G^2)$, the function
 $t \mapsto \|C^{1/2} P(t) u \|^2 $ is differentiable and
$$ \frac d {dt}  \|C^{1/2} P(t) u \|^2 = 2 Re \lan C^{1/2} P(t) u,
 C^{1/2} GP(t) u \ran. $$
 \end{lemm}
 \begin{pf}
 For each $u\in D(G)$ and each
$\lam >0$, let $v =\lam(\lam-G)^{-1}u:=\lam R(\lam, G) u$.
Obviously $v\in D(G^2)$. The inequality (\ref{2.12}) yields
\begin{eqnarray}\nonumber
 \|C^{1/2} u\|^2 &=& \frac 1 {\lam^2} \lan C^{1/2} (\lam -G) v,
 C^{1/2} (\lam -G) v \ran \\ \nonumber
 &=&\|C^{1/2} v\|^2-2\lam^{-1}Re\langle Cv,Gv\rangle +
                \lam^{-2}\|C^{1/2}Gv\|^2\\
 &\ge&(1-\lam^{-1}b)\|C^{1/2}v\|^2-a\lam^{-1}\varepsilon^{-p}\|v\|^2.\label{2.14}
\end{eqnarray}
Note that $\|u\|^2\ge\|\lam R(\lam, G) u\|^2.$ Let $\beta=\max\{b,
a\varepsilon^{-p}\}.$ It follows from (\ref{2.14}) that the
inequality
 \begin{eqnarray}
 \|C^{1/2}u\|^2+\|u\|^2&\ge&(1-\lam^{-1}b)\|C^{1/2}\lam R(\lam, G) u\|^2
+(1-\lam^{-1}a\varepsilon^{-p})\|\lam R(\lam, G)u\|^2 \nonumber \\
&\ge&(1-\lam^{-1}\beta)\Big(\|C^{1/2}\lam R(\lam, G) u\|^2 +\|\lam
R(\lam, G) u\|^2\Big). \label{3.12*}
\end{eqnarray}
The above inequality also holds for $u\in D(C^{1/2})$ since $D(G)$
is a core for $C^{1/2}$.

Note $D(C^{1/2})$ is a  Hilbert space  endowed with the graph
norm. Let $\tilde{G}$ : $ D(C^{1/2}) \to D(C^{1/2})$ be given by
 $D(\tilde{G}) = \{
u\in D(G) :  Gu \in D(C^{1/2}) \}$ and $\tilde{G}u = Gu$, for all
$u \in D(\tilde{G})$. It is easily checked that $\tilde{G}$ is
closed. Since $D(G^2)$ is a core for $G$ and $D(G)$ is a core for
$C^{1/2}$, $D(G^2)$ is a core for $C^{1/2}$ (see Lemma 2.5 of
\cite{DJ}). Thus $\tilde{G}$ is densely defined in the Hilbert
space $D(C^{1/2})$.   Let us check $R(\lambda, \tilde{G}) u =
R(\lambda, G)u$ for all $u \in D(C^{1/2})$. If $(\lambda - G) u
\in D(C^{1/2})$ for $u \in D(G)$,  then $Gu \in D(C^{1/2})$ and we
have  $ u \in D(\tilde{G})$. Since $\lambda - G$ is a bijection
from $D(G)$ to $\ch$, the range of $\lambda - \tilde{G}$ is
$D(C^{1/2})$. Thus $\lambda - \tilde{G}$ is invertible on
$D(C^{1/2})$ and $R(\lambda, \tilde{G})$ is the restriction of
$R(\lambda, G)$ to $D(C^{1/2})$.  Therefore the inequality
(\ref{3.12*}) implies that $\tilde{G}$ is the infinitesimal
generator of a strongly continuous semigroup on the Hilbert space
$D(C^{1/2})$ endowed with the graph norm. See Sec. 1 Corollary 3.8
in \cite{Paz}. This semigroup is obtained by restricting the
operators $P(t)$ to $D(C^{1/2})$. Since $D(G^2) \subset
D(\tilde{G})$, the claimed differentiation formula follows.
\end{pf}

Under assumption {\bf C} we can obtain a useful estimate of
$\mathcal{R}_{\lam, x}^{(min)} (C_\epsilon)$ where $(C_\epsilon
)_{\epsilon
>0}$ is the family of bounded regularization $C_\epsilon
=C(I+\epsilon C)^{-1}$.

\begin{prop} \label{prop2.2}
Suppose that the conditions {\bf A} and {\bf C} hold. Then, for
any $x\in (0,1)$, $\lam > max(b,1)$ and any $u \in D(G^2)$, the
bound
\begin{equation} \label{2.14-0}
(\lam -b) \sup_{\epsilon>0}\lan u, \cR^{(min)}_{\lam, x}
(C_\epsilon) u \ran \le \| C^{1/2} u \|^2 + 2 a (1-x)^{-p} \|u\|^2
\end{equation}
holds.
\end{prop}

\begin{pf} Let $(\cR_{\lambda, x}^{(n)})_{n\ge0}$ be the sequence of monotone linear maps
on $\mathcal{B}(\ch)$  defined in (\ref{3.5}). Clearly it suffices
to show that for all $n\ge0$, $\lam
>\max(b,1),\,$ $x\in(0,1)$ and $u\in D(G^2)$, the operator
$\cR_{\lambda, x}^{(n)}(C_{\epsilon})$ satisfies
\begin{equation} \label{2.14-1}
(\lam -b) \sup_{\epsilon>0}\lan u, \cR^{(n)}_{\lam, x}
(C_\epsilon) u \ran \le \| C^{1/2} u \|^2 + 2 a (1-x)^{-p}
\|u\|^2.
\end{equation}
For $n=0$, integrating by parts, we have
\begin{eqnarray}\label{2.14-2}
\lambda\lan u, R^{(0)}_{\lam, x} (C_\epsilon) u \ran & =& \lambda
\lan u, \cP_{\lam} (C_\epsilon) u \ran\nonumber\\ &=&
\lam\int_0^\infty e^{-\lam t}\|C_{\epsilon}^{1/2}P(t)
u\|^2dt\nonumber
\\ &\le& \lam \int_0^\infty e^{-\lam t} \|C^{1/2}P(t) u\|^2dt  \\
&=& \| C^{1/2} u \|^2  + 2 Re \int_0 ^\infty e^{-\lam t} \lan C
P(t) u , GP(t) u \ran dt\nonumber.
\end{eqnarray}
Two inequalities (\ref{2.12}) and (\ref{2.14-2}) yield
\begin{eqnarray} \label{2.14-3}
\lam \lan u,\cR^{(0)}_{\lam,\,x}(C_\epsilon) u\ran &\le&\| C^{1/2}
u \|^2 -(1-\varepsilon)\int_0^\infty e^{-\lam t} \|CP(t)
u\|^2dt\\&&+b\int_0^\infty e^{-\lam t} \|C^{1/2}P(t) u\|^2dt +a
\varepsilon^{-p}\int_0^\infty e^{-\lam t} \|P(t)
u\|^2dt\nonumber\\ &=&\| C^{1/2} u \|^2
-(1-\varepsilon)\int_0^\infty e^{-\lam t} \|CP(t)
u\|^2dt\nonumber\\&&+b\sup_{\epsilon>0} \lan
u,\cR^{(0)}_{\lam,\,x}(C_\epsilon) u\ran + a
\varepsilon^{-p}\int_0^\infty e^{-\lam t} \|P(t) u\|^2dt.\nonumber
\end{eqnarray}
Notice that
\begin{equation} \label{2.14-4}
\int_0^\infty e^{-\lam t} \|P(t) u\|^2dt\le \frac{1}{\lambda}
\|u\|^2.
\end{equation}
Choose $\varepsilon = 1-x$ in (\ref{2.14-3}). Then for
$\lambda>1/2$,  (\ref{2.14-1}) holds for $n=0$.

By induction, we assume that (\ref{2.14-1}) holds for an integer
$n$. It follows from (\ref{3.6-0}) and (\ref{2.14-1})  that
\begin{eqnarray}\label{2.16}
 \lan u, \cR^{(n+1)}_{\lam, x} (C_\epsilon) u \ran &=&
  \lan u, \cP_\lam (C_\epsilon) u \ran\nonumber \\&&+ x \sum_{l=1}^\infty \int_0^\infty  e^{-\lam t}\lan L_l P(t) u, \cR^{(n)}_{\lam, x} (C_\epsilon)L_l P(t)  u \ran dt\nonumber \\
& \le &\lan u, \cP_\lam (C_\epsilon) u \ran +x\frac 1{\lam -b} \sum_{l=1}^\infty \int_0^\infty e^{-\lam t}\| C^{1/2}L_l P(t)  u \|^2  dt  \nonumber \\
&& +x\frac 1{\lam -b}2a(1-x)^{-p} \sum_{l=1}^\infty \int_0^\infty
e^{-\lam t}  \|L_lP(t)u\|^2   dt.
\end{eqnarray}
By (\ref{2.1}), we have
\begin{eqnarray}\label{2.16-1}
 \sum_{l=1}^\infty \int_0^\infty e^{-\lam t}  \|L_lP(t)u\|^2
 dt&=&\int_0^\infty e^{-\lam t} \Big(-\frac{d}{dt}\|P(t)u\|^2\Big)dt\nonumber\\
  &=&\|u\|^2-\lambda\int_0^\infty e^{-\lam t}\|P(t)u\|^2 dt.
\end{eqnarray}
By (\ref{2.14-2}), we also have
\begin{eqnarray}
\nonumber && \lan u, \cP_\lam (C_\epsilon) u \ran \\
\nonumber && \quad \le \frac \lam {\lam - b}\int_0^\infty e^{-\lam
t}\|C^{1/2}P(t)u\|^2 dt - \frac b {\lam - b}\int_0^\infty e^{-\lam
t}\|C^{1/2}P(t)u\|^2 dt \\
&& \quad = \frac 1{\lam -b} \left( \|C^{1/2} u \|^2 + 2 Re
\int_0^\infty e^{-\lam t} \lan C P(t) u, GP(t) u \ran \, dt
\right) \nonumber \\&& \qquad \qquad - \frac b {\lam -
b}\int_0^\infty e^{-\lam t}\|C^{1/2}P(t)u\|^2 dt \label{**3.14}
\end{eqnarray}
 We combine  (\ref{2.16}), (\ref{2.16-1}) and  (\ref{**3.14}) to conclude that
\begin{eqnarray}\label{2.16-2}
&&(\lam -b ) \sup_{\epsilon>0}\lan u, \cR^{(n+1)}_{\lam, x}
(C_\epsilon) u\ran \le\| C^{1/2} u \|^2 \\ &&\hspace{1cm}+ 2 Re
\int_0 ^\infty e^{-\lam t} \lan C P(t) u , G P(t) u \ran dt - b
\int_0^\infty e^{-\lambda
t}\|C^{1/2}P(t)u\|^2dt\nonumber\\&&\hspace{1cm} + x
\sum_{l=1}^\infty \int_0^\infty e^{-\lam t}\| C^{1/2}L_l P(t) u
\|^2 dt+ 2a(1-x)^{-p}\Big( \|u\|^2 - \lam \int_0^\infty e^{-\lam
t}\|P(t)u\|^2 dt\Big).\nonumber
\end{eqnarray}
Next, we use (\ref{2.13}) with $\varepsilon=(1-x)/2$ to obtain
\begin{eqnarray}\label{2.16-3}
&&2xRe \int_0 ^\infty e^{-\lam t} \lan C P(t) u, G P(t) u \ran
dt+x \sum_{l=1}^\infty \int_0^\infty e^{-\lam t}\| C^{1/2}L_l P(t)
u \|^2 dt\nonumber\\ &&\quad\le \frac {x(1-x)}2\int_0 ^\infty
e^{-\lam t} \|C P(t)u\|^2dt +x b \int_0^\infty e^{-\lambda
t}\|C^{1/2}P(t)u\|^2dt \nonumber\\ &&\hspace{2cm} + 2ax(1-x)^{-p}
\int_0^\infty e^{-\lam t}\| P(t) u \|^2dt.
\end{eqnarray}
On the other hand  it follows from (\ref{2.12}) with
$\varepsilon=1/2$ that
\begin{eqnarray}\label{2.16-4}
&&2(1-x)Re \int_0 ^\infty e^{-\lam t} \lan C P(t)u,GP(t)u \ran dt\\
&&\quad\le -\frac {(1-x)} 2
\int_0^\infty e^{-\lam t}\| C P(t) u \|^2 dt\nonumber\\
&&\quad\quad+(1-x)b\int_0 ^\infty e^{-\lam t}\|C^{1/2}P(t)u\|^2dt
+ 2a(1-x) \int_0^\infty e^{-\lam t}\| P(t) u \|^2dt.\nonumber
\end{eqnarray}
Summing (\ref{2.16-3}) and (\ref{2.16-4}) yields
\begin{eqnarray}\label{2.16-5}
&&2Re \int_0 ^\infty e^{-\lam t} \lan C P(t)u, GP(t)u \ran
dt+x\sum_{l=1}^\infty \int_0^\infty e^{-\lam t}\| C^{1/2}L_l P(t)
u \|^2 dt\nonumber\\ &&\le b\int_0 ^\infty e^{-\lam
t}\|C^{1/2}P(t)u\|^2dt+2a\big(x(1-x)^{-p}+(1-x)\big)\int_0^\infty
e^{-\lam t}\|P(t) u \|^2 dt\nonumber\\ &&\le b\int_0 ^\infty
e^{-\lam t}\|C^{1/2}P(t)u\|^2dt+2a(1-x)^{-p}\int_0^\infty e^{-\lam
t}\|P(t)u\|^2dt.
\end{eqnarray}
For $\lambda>\max(b,1)$, substituting (\ref{2.16-5}) into
(\ref{2.16-2}), we obtain that
\begin{equation*}
(\lam -b) \sup_{\varepsilon>0}\lan u, \cR^{(n+1)}_{\lam, x}
(C_\epsilon) u \ran \le \| C^{1/2} u \|^2 + 2 a (1-x)^{-p}
\|u\|^2.
\end{equation*}
This completes the proof of the Proposition.
\end{pf}

\noindent {\it Proof of Theorem \ref{thm3.4}:} Let $\lam
> \max(b,1)$. Recall that for $\epsilon>0$, $C_\epsilon
=C(I+\epsilon C)^{-1}$. For $u \in D(G)$, we have
\begin{eqnarray*}
\sup_{\epsilon >0} \lan u, \cP_\lam (\Phi_\epsilon) u \ran &=& \int_0^\infty e^{-\lam t} \| \Phi^{1/2} P(t) u\|^2 dt \\
&=& \sum_{l=1}^\infty \int_0^\infty e^{-\lam t} \|L_l P(t) u\|^2 dt =\lan u, \cQ_\lam (I) u \ran.
\end{eqnarray*}
This implies that the non-decreasing family of operators
$(\cP_\lam (\Phi_\epsilon))_{\epsilon >0} $ is uniformly bounded
and since $D(G) $ is dense in $\ch$, it follows that it converges
strongly to $\cQ_\lam (I)$ as $\epsilon$ goes to 0.  By the
normality of the maps $\cQ^k_\lam$ and the equation (\ref{3.5-1}),
for any $x\in(0,1)$, we have
\begin{eqnarray*}
\sum_{k=0}^\infty x^k \lan u, \cQ_\lam^{k+1} (I) u \ran & =&
\sup_{\epsilon >0} \sum_{k=0}^\infty x^k \lan u, \cQ_\lam^k
(\cP_\lam (\Phi_\epsilon)) u \ran \\ &=& \sup_{\epsilon >0} \lan
u, \cR_{\lam, x}^{(min)} (\Phi_\epsilon) u \ran.
\end{eqnarray*}
Let $\tilde{\Phi}=\delta \Phi$.  For $\epsilon
>0$, it follows from Proposition 2.2.13 in \cite{BR} that
the bounded positive operators $\tilde{\Phi}_\epsilon$ and
$C_\epsilon$ satisfy the inequality $\tilde{\Phi}_\epsilon \le
C_\epsilon$. Applying Proposition \ref{prop2.2} we obtain the
estimate
\begin{eqnarray*}
\int_0^1  \sum_{k=0}^\infty x^{k+1}\lan u, \cQ_\lam ^{k+1} (I) u
\ran dx &=& \delta^{-1} \int_0^1 x \sup_{\epsilon >0} \lan u ,
\cR_{\lam, x}^{(min)} (\tilde{\Phi}_\epsilon )u \ran dx\\ &\le&
\delta^{-1}\int_0^1 x \sup_{\epsilon
>0} \lan u, \cR_{\lam,x}^{(min)} (C_\epsilon) u \ran dx\\ &\le& \delta^{-1}\int_0^1 x
(\lam -b )^{-1} \Big( \|C^{1/2}u \|^2 + 2a (1-x) ^{-p} \|u\|^2
\Big) dx\\[0.2cm]
&<&\infty.
\end{eqnarray*}
By (\ref{3.3}) and Lemma \ref{lemm2.1} we have
$s-\lim_{n\rightarrow\infty}\mathcal{Q}_\lam^n(I)=0$, which
implies that the minimal q.d.s. is conservative. $\quad \square$.

\section{Applications}
\quad In this section we obtain some relative bounds to apply our
sufficient conservativity condition of Theorem \ref{thm3.4} to a
concrete example.

 Let $\ch=L^2(\R^n,dx)$ and $W:\R^n\rightarrow\R$ be the real
 valued function. We are looking for the condition that there exist
constants $a>0$ and $p<1$ such that $$\|W \varphi\|^2\le
\varepsilon\|-\Delta \varphi\|^2+a\varepsilon^{-p}\|
\varphi\|^2,\quad \varphi\in C_0^2(\R^n)$$ holds for any
$\varepsilon>0$, where $\Delta$ is a Laplacian operator and
$C_0^2(\R^n)$ is the set of twice continuously differentiable
functions with compact support on $\R^n$. We prove first the
following :
\begin{lemm} \label{lemm3.1}
For a given $n\in\mathbb{N}$, let $\alpha$ be a nonnegative real
number satisfying  $n/(1+\alpha)<2$. If $W\in
L^{2+2\alpha}(\R^n),$ there exist $a>0$ and $p<1$ such that the
bound $$ \|W \varphi\|^2\le\varepsilon\|-\Delta
 \varphi\|^2+a\varepsilon^{-p}\| \varphi\|^2
$$ holds for any $\varepsilon>0$ and $ \varphi\in D(-\Delta)$.
\end{lemm}
\begin{pf}
Since $C_0^\infty(\R^n)$, the space of infinitely differentiable
functions with compact support, is a core for $-\Delta$, it is
sufficient to show the bound for any $\varphi\in
C_0^\infty(\R^n)$.

We use the method employed in the proof of  Theorem IX 28 in
\cite{RS}. Assume $W^{1+\alpha}\in L^2(\R^n)$. Denote by $\hat{f}$
the Fourier transform of $f\in\ch$. For $\varphi\in
C_0^\infty(\R^n)$, we have
\begin{eqnarray}\label{4.1}
\|W^{1+\alpha}\varphi\|_2^2&\le&\|W^{1+\alpha}\|_2^2\|\varphi\|_\infty^2,\\
\|\varphi\|_\infty&\le&(2\pi)^{-n/2}\|\hat{\varphi}\|_1\nonumber
\end{eqnarray}
and \begin{equation}\label{4.1-1} \|\hat{\varphi}\|_1^2\le
C\|({\lambda}^4+1)^{(1+\alpha)/2}\hat{\varphi}\|_2^2,
\end{equation} where
$C=\|({\lambda}^4+1)^{-(1+\alpha)/2}\|^2_2<\infty$ since
$\alpha>\frac{n}{2}-1.$

For any $r>0$, let
$\hat{\varphi}_r(\lambda)=r^n\hat{\varphi}(r\lambda)$. Then
\begin{eqnarray*}\label{4.2}
\|\hat{\varphi}_r\|_1&=&\|\hat{\varphi}\|_1,\\
\|({\lambda}^4+1)^{(1+\alpha)/2}\hat{\varphi}_r\|_2^2&=&
\int_{\R^n}({\lambda}^4+1)^{1+\alpha}r^{2n}
|\hat{\varphi}(r\lambda)|^2d^n\lambda\nonumber\\&=&
r^n\|(r^{-4}{\lambda}^4+1)^{(1+\alpha)/2}\hat{\varphi}\|^2_2.
\end{eqnarray*}
Thus using (\ref{4.1-1}) for $\hat{\varphi}_r$, and these
equalities, we obtain
\begin{equation}\label{4.2-1}
\|\hat{\varphi}\|_1^2\le
Cr^n\|(r^{-4}{\lambda}^4+1)^{(1+\alpha)/2}\hat{\varphi}\|^2_2.
\end{equation}
Substituting (\ref{4.2-1}) into (\ref{4.1}), by Plancherel's
Theorem,  there is a constant $C_1>0$ such that
\begin{equation*}
\|W^{1+\alpha}\varphi\|_2^2\le C_1
r^n\|(r^{-4}{\Delta}^2+1)^{(1+\alpha)/2}\varphi\|^2_2,
\end{equation*}
which implies
\begin{equation}\label{4.5}
W^{2+2\alpha}\le C_1r^n(r^{-4}{\Delta}^2+1)^{1+\alpha}.
\end{equation}
Suppose that $A$ and $B$ are self-adjoint operators such that
$$0\le B\le A.$$ Then the above implies that $$ 0\le B^t\le A^t$$
for any $t\in [0,1]$( see Problem 51 of Chapter VIII of \cite{RS}
and also the Heinz-Kato theorem in \S 2.3.3. of \cite{Ta}). Thus
we have
 $$ W^2\le
C_2r^{n/{(1+\alpha)}}(r^{-4}{\Delta}^2+1),$$
 which yields
 $$\|W\varphi\|^2\le C_2r^{-(4-n/{(1+\alpha))}}\|-\Delta \varphi\|^2 +
 C_2r^{n/{(1+\alpha)}}\|\varphi\|^2.$$
 Choose
$\varepsilon=C_2r^{-(4-n/{(1+\alpha)})}$.  Then we obtain
 $$\|W\varphi\|^2\le \varepsilon\|-\Delta \varphi\|^2 +
 a\varepsilon^{-p}\|\varphi\|^2$$  where $p=n(4(1+\alpha)-n)^{-1}$. Since
 $n/(1+\alpha)<2$, $p<1$.
 If we choose $r$ large enough, the
bound follows.
\end{pf}
\begin{rem}\label{remm3.2} (a) In Lemma \ref{lemm3.1},
 one can choose $\alpha=0$ for $n=1$. Notice that $\alpha>0$ for
$n=2$ and $\alpha>1/2$ for $n=3$, etc.

(b) Let the dimension $n=1,2,3$. If $W\in L^4(\mathbb{R}^n, dx)$,
than $W^2\in L^2(\mathbb{R}^n, dx)$ and so $W^2$ is relatively
bounded by $-\Delta$ (see Theorem X.15 of \cite{RS}). Thus $W^2$
is relatively form bounded by $-\Delta$, i.e.,
$$
\| W\varphi \|^2 \le b \lan \varphi,(-\Delta +1) \varphi \ran ,
\quad \varphi \in C_0^\infty (\R^n),
$$
where $b$ is a constant. See also Theorem X.18 (b) of \cite{RS}.
\end{rem}

  In the rest of this section, we apply Theorem \ref{thm3.4} and
  Lemma \ref{lemm3.1} to a model of heavy ion collision proposed
  by Alicki \cite{Al}.

\vspace{0.5cm} \noindent
 {\bf {Example 4.1}}
(Q.d.s. in a model for heavy ion collision)

\noindent
  Let $\ch=L^2(\R^3)$. We denote by $\partial_k=\frac{\partial}{\partial
  x_k}(\,k=1,2,3)$ differential operators with respect to the $k$ th coordinate
and $\partial_{lk}=\frac{\partial^2}{\partial x_k\partial
x_l}(\,l,k=1,2,3)$. For any measurable function $T$, we denote the
(distributional) derivative  $\frac{\partial T}{\partial x_l}$ by
$(T)_l$, $\,l=1,2,3$. Consider the operators $L_l,$ for $l=1,2,3$
 \begin{eqnarray}
  L_l u &= &w(x_l + \alpha \partial_l) u, \label{L4.4} \\
 D(L_l ) &=&\{ u \in L^2(\R^3) :\, \text{the distribution } \,\,L_l u \in
 L^2(\R^3)\} \nonumber
 \end{eqnarray}
 where  $w, \alpha \in \R$ are non-zero real constants, and $L_l=0$ for $l\ge 4$.
 Let $V$ be a real measurable function.  Consider the operators $H$ and $G$ given by
 \begin{eqnarray}
 H u &=& (-\frac1{2} \Delta +V )\, u, \label{4.6n}\\
 G u &=& -i H u-\frac12\sum_{l=1}^\infty L_l^*L_l u \nonumber
 \end{eqnarray}
for $u \in C_0^\infty (\R^3)$. Let us assume that the following
properties hold:
\begin{enumerate}
\item[(1)] $w^2 \alpha^2 \ge 2$
\item[(2)] $|V(x) | \le \frac14 w^2 (x^2 + b_1) $ for some
constant $b_1 >0$, where $x^2 = x_1^2 + x_2^2 + x_3^2$.
\item[(3)] There exist real measurable functions $U_1$ and $U_2$ and
positive constants $b_2$, $b_3$ such that $U_1 \in L^\beta (\R^3)$
for some $\beta >3$, $U_2 (x) \le b_2 (|x| + b_3)$ and the bounds
\begin{equation} \label{4.7n}
|(V)_l| \le U_1 + U_2
\end{equation}
hold for $l=1,2,3.$
\end{enumerate}

 For an instance the function $V(x) = \frac 14 w^2 | x|^\nu$, $0< \nu \le 2$, satisfies the conditions
 (2) and (3) in the above.  Let us mention that in the
example proposed by Alicki \cite{Al}, the constant $w$ in
(\ref{L4.4}) is a function $W(x)$ proportional to
$\sqrt{\gamma(x)}$ where $\gamma (x)$ represents a friction force.
The conservativity of this q.d.s. has been already investigated in
\cite{CF2} under appropriate (boundedness) assumptions on $V,W$
and their derivatives.   In this paper we only consider the case
that $W(x)$ is a constant to avoid unnecessary notational
complications involved.

We apply  Theorem \ref{thm3.4} and
  Lemma \ref{lemm3.1} to show that the minimal q.d.s. constructed from
above operators $L_l$ and $G$ given in (\ref{L4.4}) and
(\ref{4.6n}) respectively is conservative. We will check that the
main inequalities (\ref{2.12}) and (\ref{2.13}) hold for $u \in
C_0^\infty (\R^3)$. The most difficult problem  is to extend the
inequalities  to every $u \in D(G^2)$. In order to overcome this
problem, we need technical estimates.

\begin{lemm}\label{lemm*4.2*}
For all $u \in C_0^\infty(\R^3)$, the bounds
\begin{equation} \label{4.8n}
\langle u , (\alpha^4 \Delta^2 + x^4) u \rangle \le \langle u,
(-\alpha^2 \Delta + x^2 +3|\alpha| )^2 u \rangle
\end{equation}
and
\begin{equation} \label{4.9n}
\| \frac 12 w^2 (-\alpha^2 \Delta + x^2 -3 \alpha ) u \|^2 \le b_4
\|G u \|^2 + b_5 \| u\|^2
\end{equation}
for some $b_4 >1$ and $b_5 >0$ hold.
\end{lemm}
\begin{pf} A direct computation shows that
\begin{eqnarray*}
(-\alpha^2 \Delta + x^2)^2 &=& \alpha^4 \Delta^2 + x^4 - \alpha^2
(\Delta x^2 + x^2 \Delta) \\
&=& \alpha^4 \Delta^2 + x^4 - \alpha^2 (2\sum_{k=1}^3 \partial_k
x^2 \partial_k +6) \\
&\ge& \alpha^4 \Delta^2 + x^4 -6\alpha^2,
\end{eqnarray*}
as a bilinear form on the domain $C_0^\infty(\R^3)$. This proves
the bound (\ref{4.8n}).

Next we prove the bound (\ref{4.9n}). Put  \begin{eqnarray}
 G_0&=&  -\frac12 \sum_{l=1}^3
L_l^*L_l \label{4.10n} \\
&=& -\frac 12 w^2 (-\alpha^2 \Delta + x^2 -3\alpha). \nonumber
\end{eqnarray}
 We have that as bilinear
forms on $C_0^\infty (\R^3)$
\begin{eqnarray}
G^* G &=& (i H + G_0)(-i H + G_0) \nonumber \\
&=& H^2 + G_0 ^2 + i [H, G_0] \nonumber \\
& \ge & G_0^2 + i [H, G_0 ], \label{4.11n}
\end{eqnarray}
and
\begin{eqnarray*}
i[H,G_0 ] &=& \frac{iw^2} 4 [\Delta, x^2 ] + \frac{iw^2\alpha^2} 2
[V,
\Delta] \\
&=& \frac{iw^2} 2  \sum_{l=1}^3 (\partial_l x_l + x_l
\partial_l) -\frac{iw^2\alpha^2} 2  \sum_{l=1}^3 \big(\partial_l (V)_l + (V)_l
\partial_l \big) \\
&\ge &- \frac {w^2} 2  (-\Delta + x^2) - \frac {w^2 \alpha^2} 2
(-\Delta + \sum_{l=1}^3 | (V)_l|^2 ).
\end{eqnarray*}
It follows from (\ref{4.7n}) that
$$
i[H, G_0 ] \ge -\frac {w^2} 2 (1+\alpha^2) (-\Delta +x^2) -
3w^2\alpha^2 (U_1^2 + U_2^2).
$$
The bound (\ref{4.8n}) implies that $(-\Delta +x^2 )^{1/2}$ is
infinitesimally small with respect to $G_0$. By the condition (3)
and Lemma \ref{lemm3.1} (Remark \ref{remm3.2} (a)), $U_1$ and
$U_2$ are also infinitesimally small with respect to $G_0$. Thus
there exist constants $ 0<a<1$ and $b >0$ such that
$$
i[H, G_0] \ge -a G_0^2 - b
$$
as a bilinear form on $C_0^\infty (\R^3)$. The bound (\ref{4.9n})
follows from (\ref{4.11n}) and the above bound.
\end{pf}

Recall that
$$
G = -i (-\frac 12 \Delta + V) + G_0
$$
where $G_0$ is given as (\ref{4.10n}). Notice that $G_0$ is
essentially self-adjoint on $C_0^\infty (\R^3)$. Since
$w^2\alpha^2 \ge 2$ by the condition (1), the bound (\ref{4.8n})
implies that $-\frac12 \Delta$ is $G_0$-bounded with relative
bound smaller than equal to $1/2$. The condition (2) and the bound
(\ref{4.8n}) imply that $V$ is $G_0$-bounded with relative bound
smaller than 1/2. Thus $-iH$ is relatively bounded perturbation of
$G_0$ with relative bound smaller than 1. Thus Assumption {\bf A}
holds.

 We show that the minimal q.d.s. is conservative applying Theorem
 \ref{thm3.4}. Let us choose the operator $C$,
 \begin{equation}\label{C4.7}
 C=  w^2 (-\alpha^2 \Delta + x^2 + 3|\alpha|) =
 \sum_{l=1}^3 L_l^* L_l + b_6 = -2 G_0 + b_6,
 \end{equation}
 $$
 D(C) =\{ u \in L^2(\R^3) | \,\text{ the distribution}\,\, Cu \in
 L^2(\R^3) \}
 $$
 where $b_6 = 3 w^2 (|\alpha|-\alpha)$. Using the relation (\ref{4.9n}) and the fact that
 $-iH$ is relatively bounded perturbation of $G_0$, we obtain  that $G$ and $C$ are relatively bounded
 with respect to each other and so $D(G)$ = $D(C)$.

 We will check that the operator $C$ satisfies the assumption {\bf
 C}. Hypothesis (a) and (b) are trivially fulfilled.
 Now we will check (c). First, we have that as bilinear forms on
 $C_0^\infty (\R^3)$,
 \begin{eqnarray}
 [C, -\frac12 \Delta + V] &=& -\alpha^2 w^2 [\Delta, V] -\frac 12 w^2
 [x^2, \Delta] \nonumber \\
 &=& -\alpha^2 w^2 \sum_{l=1}^3 \big( \partial_l (V)_l + (V)_l
 \partial_l \big) + w^2 \sum_{l=1}^3 \big( \partial_l x_l + x_l
 \partial_l ) \nonumber \\
 &\le& \alpha^2 w^2 \big(-\Delta + \sum_{l=1}^3 (V)_l^2 \big) + w^2
 (-\Delta +x^2), \label{*1.1n}
 \end{eqnarray}
 and
 \begin{eqnarray*}
 [C, L_l ] &=& w^3 \big( -\alpha^2 [\Delta, x_l ] + \alpha [x^2 ,
 \partial_l] \,\big) \\
 &=& -2 w^3 \alpha (\alpha \partial_l + x_l) =-2 w^2 \alpha L_l,
 \end{eqnarray*}
and so
\begin{equation}\label{*1.2n}
\sum_{l=1}^3 L_l^* [C, L_l] = -2w^2\alpha \sum_{l=1}^3 L_l^* L_l =
-2w^2\alpha C + b_6.
\end{equation}
By direct computation, we have
$$
CG +G^* C + C^2 = -i[C, -\frac12 \Delta +V] + b_6 C,
$$
and
\begin{eqnarray*}
&& C G+ G^* C + \sum_{l=1}^3 L_l^* C L_l \\
&& \quad = -i[C, -\frac 12 \Delta + V] +\frac 12 \sum_{l=1}^3
\big( L_l^* [C, L_l] + (L_l^*[C, L_l])^* \big) \\
&& \quad = -i [C, -\frac12 \Delta + V] -2w^2 \alpha C + b_6,
\end{eqnarray*}
as bilinear forms on $C_0^\infty (\R^3).$ Substituting
(\ref{*1.1n}) and (\ref{*1.2n}) into the above equations, and
using the fact that $-\Delta, -\Delta+ x^2$ are relatively form
bounded with respect to $C$, we have that for $u \in C_0^\infty
(\R^3)$,
\begin{equation} \label{e4.9}
2 Re \langle Cu , G u \rangle + \| Cu\|^2 \le b_7 \langle u , Cu
\rangle + \alpha^2 w^2 \sum_{l=1}^3 \| (V)_l u \|^2,
\end{equation}
and
\begin{eqnarray}
&& 2 Re \langle Cu , Gu\rangle + \sum_{l=1}^3 \langle L_l u , C
L_l u
\rangle \nonumber \\
 && \quad \qquad\le b_8 \langle u, Cu
\big\rangle + \alpha^2 w^2 \sum_{l=1}^3 \| (V)_l u \|^2
\label{e4.10},
\end{eqnarray}
 where $b_7, b_8 >0$.

 Note $|(V)_l| \le U_1 + U_2$ for $l=1,2,3$ with $U_1  \in L^\infty (\R^3)$  where $\beta>3$, and
  $U_2(x) \le b_2 (|x| + b_3)$. Applying  Lemma \ref{lemm3.1} to
(\ref{e4.9}) and (\ref{e4.10}), we obtain (\ref{2.12}) and
(\ref{2.13}) for $u \in C_0^\infty(\R^3)$.

We want to extend the inequality (\ref{2.12}) and (\ref{2.13}) to
the domain $D(G)$. For $u\in D(G)$, there exists a sequence $\{u_n
\}$ of elements of $C_0^\infty (\R^3)$ such that
$$ \lim_{n \to \infty} u_n =u, \,\,\lim_{n \to \infty} Cu_n =Cu,
\,\,\lim_{n \to \infty} Gu_n =Gu,
$$
by the relation (\ref{4.9n}). Then the relation (\ref{2.12}) holds
for $u\in D(G)$. Also the relation (\ref{2.13}) implies that $ \{
C^{1/2} L_l u_n \}_{n\ge 1}$ is a Cauchy sequence. Therefore it is
convergent and it is easy to deduce that (\ref{2.13}) holds for $u
\in D(G)$.

Recall that $\Phi= \sum_{l=1}^3 L_l^* L_l $ and $C= \sum_{l=1}^3
L_l^* L_l + b_6$. Hence the conditions of Theorem
   \ref{thm3.4} also hold and the minimal
q.d.s. is conservative.

\begin{rem} \label{rem4.2}
 Let us remind the condition of derivatives of  $V$,
 $|(V)_l| \le U_1 + U_2$ for $l=1,2,3$.
    One can use the previous criterion in
\cite{CF2} to show the conservativity for $ U_1 \in  L^4(\R^3)$
(see Remark \ref{remm3.2} (b)).  Applying our result, we extend
the range of $(V)_l$, i.e., $U_1
 \in L^\beta(\R^3) $ where $\beta >3$.
\end{rem}

\vspace*{0.2cm} \noindent {\bf Acknowledgement} : This work was
supported by Korea Research Foundation Grant
(KRF-2003-005-C00010).

\end{document}